\documentclass{iau} 
\usepackage{graphicx}
\newcommand{\Mrate}{\dot{M}}
\newcommand{\Mdot}{\mbox{\,$\rm M_{\odot}$}}        
\newcommand{\Ldot}{\mbox{\,$\rm L_{\odot}$}  }        
\newcommand{\kms}{\,km s$^{-1}$}   			
\newcommand{\aov}{$\alpha_{\rm ov}$}		

\title[Massive star evolution revealed in the Mass-Luminosity plane] 
{Massive star evolution revealed in the Mass-Luminosity plane}

\author[Erin R. Higgins \& Jorick S. Vink]   
{Erin R. Higgins$^1$$^2$$^3$$^4$ \& Jorick S. Vink$^1$$^4$}

\affiliation{$^1$Armagh Observatory and Planetarium, College Hill, Armagh BT61 9DG, N. Ireland \\[\affilskip] $^2$Queen's University of Belfast, Belfast BT7 1NN, N. Ireland \\[\affilskip] $^3$Dublin Institute for Advanced Studies, 31 Fitzwilliam Place, Dublin, Ireland \\[\affilskip] $^4$Kavli Institute for Theoretical Physics, University of California, Santa Barbara, CA 93106, USA \\email{: \tt eh@arm.ac.uk; jsv@arm.ac.uk}}

\pubyear{2019}
\volume{346} 
\jname{High-mass X-ray binaries: illuminating the passage from massive binaries to merging compact objects}
\editors{A.C. Editor, B.D. Editor \& C.E. Editor, eds.}
\begin{document}

\maketitle

\begin{abstract}
Massive star evolution is dominated by key physical processes such as mass loss, convection and rotation, yet these effects are poorly constrained, even on the main sequence. We utilise a detached, eclipsing binary HD166734 as a testbed for single star evolution to calibrate new MESA stellar evolution grids. We introduce a novel method of comparing theoretical models with observations in the 'Mass-Luminosity Plane', as an equivalent to the HRD (see Higgins \& Vink, 2018). We reproduce stellar parameters and abundances of HD166734 with enhanced overshooting (\aov$=$0.5), mass loss and rotational mixing. When comparing the constraints of our testbed to the systematic grid of models we find that a higher value of \aov $=$ 0.5 (rather than \aov $=$ 0.1) results in a solution which is more likely to evolve to a neutron star than a black hole, due to a lower value of the compactness parameter.
\end{abstract}

\firstsection 

\section{Introduction}
The lives of massive stars (M$_{\rm init}$ $\geq$ 8\Mdot) are influenced by the key processes acting on their structure, including stellar winds, convection and rotation. The degree at which these processes effect evolutionary paths is dictated by the initial mass, metallicity and multiplicity of the star. In the mass range of $\sim$8-30\Mdot~ rotation is the dominant effect, whereas above $\sim$30\Mdot~the evolution is heavily dominated by mass loss via stellar winds, \cite[e.g. Maeder \& Meynet (2000)]{MM00}. 
The main-sequence (MS) lifetime is dependent on the extension of the convective core by overshooting due to extra mixing of hydrogen (H) in the core, though this processes remains largely unresolved it is key in determining the final phases of evolution (e.g. final masses which form neutron stars or black holes) since the MS accounts for 90\% of O star lifetimes, \cite[(Claret \& Torres, 2017)]{ClaretTorres}. \cite[Martins \& Palacios (2013)]{Martins13} present an overview of current evolutionary codes \cite[e.g. Ekstr\"{o}m et al. (2012), Brott et al. (2011b)]{Gen12, Bonn11} as well as their implementations of various physical processes. The differing treatment of rotation, mass loss and convective overshooting has led to a diversity of evolutionary masses and MS-lifetimes. \cite[Weidner \& Vink (2010)]{WeidnerVink} discuss the effects this may have on the mass determination of O stars and further discrepancies when compared to spectroscopic masses (i.e. the 'mass discrepancy' problem). In this paper we aim to compare new and existing prescriptions with one code, utilising the highly flexible MESA code \cite[e.g. Paxton et al. 2011]{Paxton11}.
We investigate the evolution of a high mass, detached, eclipsing binary HD166734, modelled as a testbed for single star evolution with constraints on \aov\ and $\Mrate$. Dynamical masses of 39.5\Mdot~ and 33.5\Mdot, for the primary and secondary respectively, have allowed exploration of the dominant processes for the entire O star mass range, since in this mass regime the effects of rotation, mass loss and overshooting all play as role as they interact and overlap. We develop a novel method of analysing the evolution of our models alongside observations of HD166734 in the Mass-Luminosity plane, and present a calibrated grid of rotating and non-rotating models with a sample of Galactic O stars.

\section{Methodology}
The stellar evolution code MESA has been utilised in this study in developing a set of theoretical models for calibration of physical processes such as convective overshooting, mass loss and rotation. The convective core boundary is defined by the Ledoux criterion with step-overshooting applied as the extension of the core by a factor \aov~ of the pressure scale height H$_{\rm p}$. Mass-loss rates are employed by the \cite[Vink et al. (2001)]{Vink01} prescription, accounting for the bi-stability jump at 21kK and metallicity dependancies. We adopt the default metallicity of Z=0.02 in MESA to provide comparisons with a galactic observations. Rotation is applied in a fully diffusive approach with Eddington-Sweet circulation and dynamical and secular shear instabilities accounted for.

\cite[Mahy et al. (2017)]{Mahy2017} provide a rare opportunity in the analysis the non-interacting binary HD166734, since the dynamical masses of both stars are in excellent agreement with their spectroscopic masses, thus allowing calibration of evolutionary masses through our models. Table \ref{HD166734} highlights the main stellar parameters of HD166734, including exact positions in the HRD and surface nitrogen abundances. We apply an equal-age assumption for the binary such that they have evolved from the same initial stage, allowing for constraints of the MS-width and rotation rates.

\begin{table}
\caption{\label{HD166734}HD\,166734 Properties}
\centering
\begin{tabular}{lcc}
\hline\hline
 & Primary & Secondary\\ 
\hline
$T_{\rm eff}$ [K] & 32000 $\pm$1000 & 30500 $\pm$1000\\
log(L/\Ldot) & 5.840 $\pm$ 0.092 & 5.732 $\pm$ 0.104\\
$M_{\rm dyn}$ [\Mdot] & 39.5 $\pm$ 5.4 & 33.5 $\pm$ 4.6\\
$M_{\rm spec}$ [\Mdot] & 37.7 $\pm$ 29.2 & 31.8 $\pm$ 26.6\\
v sin i [\kms] & 95 $\pm$ 10 & 98 $\pm$ 10\\
$\rm [N/H]$  & 8.785 & 8.255\\
\hline
\end{tabular}
\caption{Fundamental properties of HD\,166734, from \cite[Mahy et al. (2017)]{Mahy2017}.}
\end{table}

\section{Mixing and Mass loss}

\begin{figure}
\centering
	\includegraphics[width = 8cm]{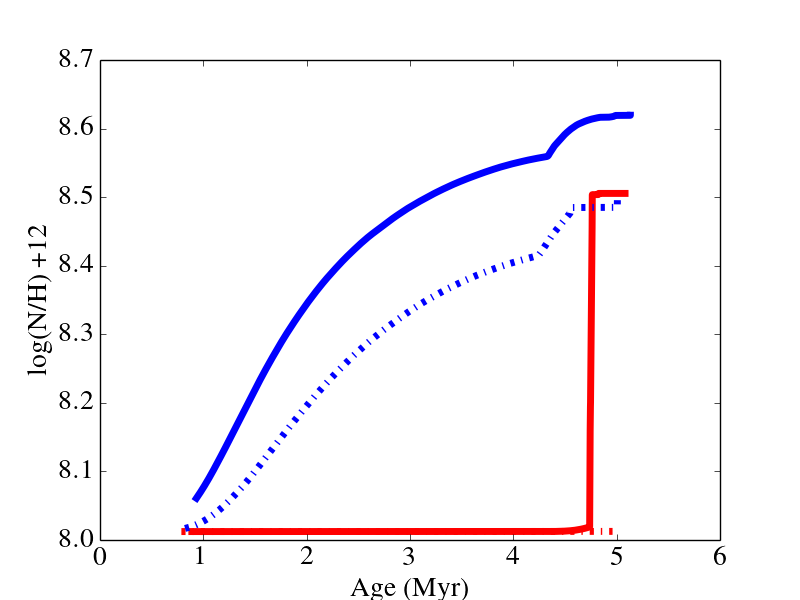}
	\caption{\footnotesize Surface nitrogen abundances as a function of stellar age for varying \aov. The blue lines represent rotating, 40\Mdot~ models with an initial rotation rate of 200\kms, \aov=0.1 (dash-dotted), and \aov=0.5 (solid). Red lines show the corresponding non-rotating models for the same mass and values of \aov\ respectively.}
	\label{N/H}
\end{figure}

In order to minimise interacting processes, we firstly calculated non-rotating models which solely employ mass loss and overshooting as mixing processes, with the aim of reproducing the 3:1 ratio of surface nitrogen abundances between the primary and secondary, and HRD positions simultaneously. We explore a range of factors of the \cite[Vink et al. (2001)]{Vink01} mass-loss prescription (0.1-10 times), alongside a variety of overshooting \aov~ = 0.1-0.8, and find that all variations of this parameter space result in either negligible surface nitrogen enrichment, or a factor of 10 enrichment, representative of CN-equilibrium. We conclude that a large value of \aov~ coupled with an increased mass-loss rate results in stripping of the stellar envelope exposing CNO material at the core boundary. Thus another mechanism of mixing is required to reproduce any intermediate surface enrichment, we hence explore the addition of rotation to our models. Figure 1 demonstrates the surface nitrogen abundances of both non-rotating (red) and rotating (blue) models for a range of \aov\ and $\Mrate$, we note that intermediate enrichment only occurs for rotating models. However, when analysing these models we were unable to simultaneously reproduce the stellar parameters in a HRD for both primary and secondary due to the dependance of rotation on mass loss. We found that only by removing the default implementation of rotationally-induced mass loss could we disentangle each physical process to reach a solution in the HRD.
\section{The Mass-Luminosity Plane}
Observations of HD166734 suggest high initial masses representative of the observed luminosities, yet present challenges when comparing this to the much lower dynamical masses. We find that an extreme factor of mass-loss rate would be required to reproduce both the observed luminosities and dynamical masses, though in this case factors 2 and above result in a significant drop in luminosity such that the observed luminosity cannot be reached. 

We hence developed a method of simultaneously reproducing the mass and luminosity of our observations via a novel tool termed here as the 'mass-luminosity plane' (M-L plane). Figure 2 illustrates an evolutionary model beginning the ZAMS at the red dot and evolving along the vector with time, or temperature as in the HRD. We find that at a given evolutionary stage or observed temperature, we may calculate the length of this vector with respect to initial mass and luminosity setting the initial position, and an observed temperature setting the final position or current evolutionary stage. We thus can use this vector length to constrain physical processes in our theoretical models to reproduce observations provided by the testbed HD166734. We observe that the length (from initial - final positions) can only be \textit{further} extended by rotation or overshooting, while the gradient of the vector relies on the multiplication factor of the mass-loss rate. 

We exclude extreme factors of the \cite[Vink et al. (2001)]{Vink01} mass-loss rate since a factor of 0.1 results in an almost vertical vector which cannot reproduce dynamical masses, as well as factors beyond 1.5 times the \cite[Vink et al. (2001)]{Vink01} prescription since this results in a much too shallow gradient to reproduce observed luminosities, leaving an accepted parameter range of 0.5-1.5 time the \cite[Vink et al. (2001)]{Vink01} prescription. When constraining our models in the M-L plane with observations of HD166734 we found that enhanced overshooting was required to reproduce stellar parameters simultaneously (\aov=0.3 $\pm$0.1 and 0.5 $\pm$ 0.1 for the primary and secondary respectively), since initial rotation rates of 250\kms\ and 120\kms\ for the primary and secondary respectively were also calibrated via surface nitrogen enrichments, and extra mixing was still required.

\begin{figure}
\centering
\includegraphics[width = 8cm]{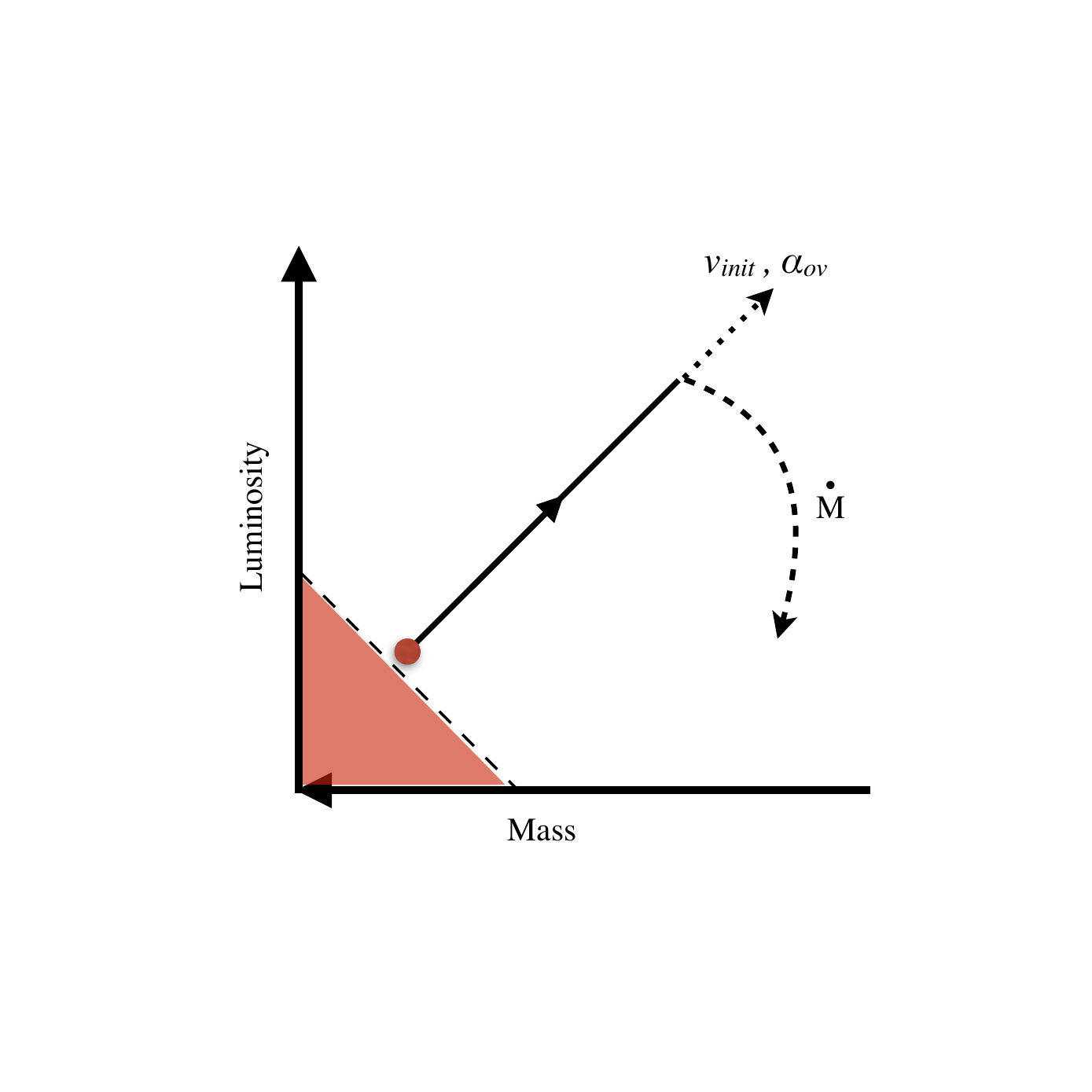}
\caption{\footnotesize Illustration of the Mass-Luminosity plane with a typical evolutionary track entering the ZAMS at the red dot, evolving along the black arrow. The dotted vector suggest how increased rotation and/or convective overshooting may extend the M-L vector. The curved dashed line represents the gradient at which mass-loss rates effect this M-L vector. The red solid region represents the boundary set by the mass-luminosity relationship, and as such is forbidden. }
\label{Cartoon}
\end{figure}

\section{Grid analysis}
We consolidate our results for HD166734 by comparing our constraints to a galactic sample of 30 O stars from \cite[Markova et al. (2018)]{Markova}. We calculated a systematic grid of models for initial masses 8-60\Mdot, initial rotation rates of 0-500\kms~ and for two extreme values of \aov $=$ 0.1 and 0.5, see figure 3 for example. We identify the main calibrations from this study as extended overshooting to \aov $=$ 0.5, and reduced mass loss with rotation by excluding rotationally-induced mass loss. We note that comparing models of \aov $=$ 0.1 and 0.5, we find that the luminosity cut-off for red supergiant evolution is much higher ($\sim$ log L/\Ldot~$=$6.0) for \aov$=$0.1, than for \aov$=$0.5 ($\sim$ log L/\Ldot~$=$5.5-5.8), regardless of rotation rate or mass-loss factor. We also note that enlarging the overshooting region may have consequences for the compactness parameter ($\zeta_{2.5}$) in the final phases of our models. We find a lower value of $\zeta_{2.5}$ with an enhanced overshooting of \aov$=$0.5 (when compared to \aov$=$0.1), thus making it easier to create neutron stars and more challenging to evolve to black holes.

\begin{figure}
\centering
\begin{minipage}[b]{0.475\linewidth}
  \includegraphics[width = 6cm]{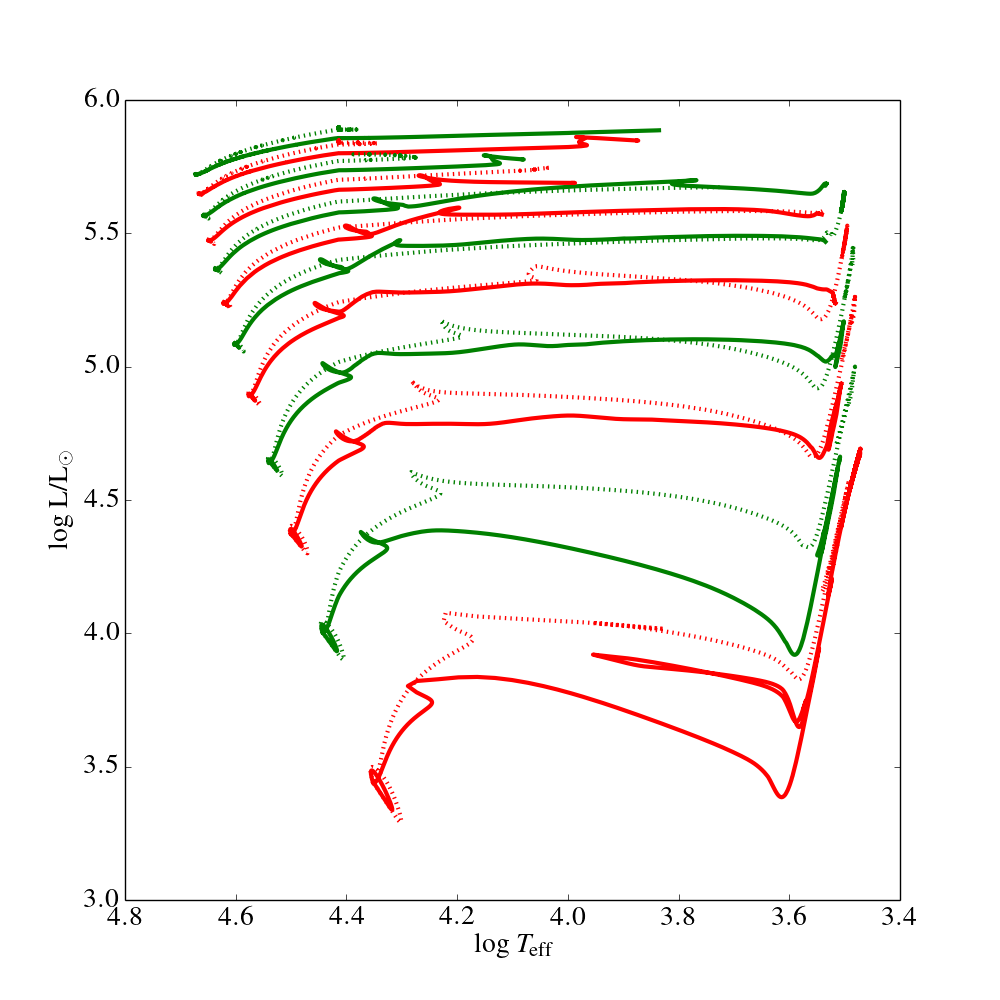}
  \caption{\footnotesize Grid results for the non-rotating models of mass range 8 - 60\Mdot employing \aov $=$ 0.1 (solid lines) and  \aov $=$ 0.5 (dashed lines).}
  \label{grid0}
\end{minipage}
\quad
\begin{minipage}[b]{0.475\linewidth}
  \includegraphics[width = 6cm]{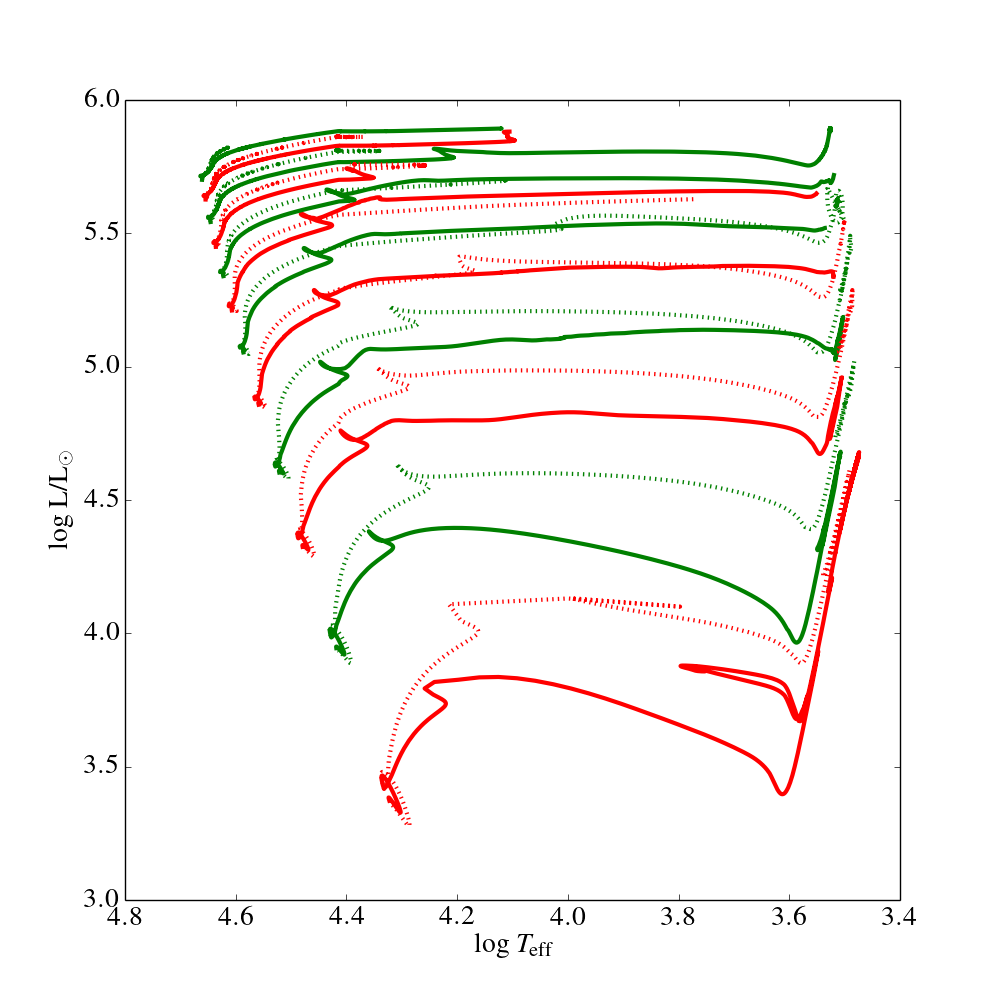}
    \caption{\footnotesize Grid results for the rotating models with initial rotation rates of 300\kms for the mass range 8 - 60\Mdot\ employing \aov $=$ 0.1 (solid lines) and  \aov $=$ 0.5 (dashed lines).}
    \label{grid3}
\end{minipage}
\end{figure}

\section{Discussion and Conclusion}
We present a novel method of comparing evolutionary tracks with observations in the Mass-Luminosity plane (Higgins \& Vink, 2018), demonstrated by our constraints of the testbed, detached binary HD166734. We overlay these calibrations with a systematic grid of models alongside a galactic sample of O stars, which reveals dependancies of \aov~ and $\Mrate$ on the final stages of evolution. We emphasise the necessity of rotational mixing in stellar models for reproducing observed intermediate surface enrichments. 

We conclude that extra mixing by overshooting is required to simultaneously reproduce the observed luminosities and dynamical masses of HD166734 in the M-L plane, while we exclude extreme factors of the mass-loss rate that lie beyond 0.5-1.5 times the \cite[Vink et al. (2001)]{Vink01} rates. Finally, we omit the application of rotationally-induced mass loss since stellar parameters of HD166734 cannot be simultaneously reproduced with this theory.

\end{document}